# Medical Simulation and Training: "Haptic" Liver

**Felix G. Hamza-Lup[1], Adrian Seitan[2],
Dorin M. Popovici[2], Crenguta M. Bogdan[2]**

(1) Computer Science and Information Technology
Armstrong Atlantic State University, Savannah, USA
(2) Mathematics and Informatics
Ovidius University, Constanta, Romania
E-mail: Felix.Hamza-Lup@armstrong.edu

**Abstract**

*Tactile perception plays an important role in medical simulation and training, specifically in surgery. The surgeon must feel organic tissue hardness, evaluate anatomical structures, measure tissue properties, and apply appropriate force control actions for safe tissue manipulation. Development of novel cost effective haptic-based simulators and their introduction in the minimally invasive surgery learning cycle can absorb the learning curve for residents. Receiving pre-training in a core set of surgical skills can reduce skill acquisition time and risks.*

*We present the development of a cost-effective visuo-haptic simulator for the liver tissue, designed to improve practice-based education in minimally invasive surgery. Such systems can positively affect the next generations of learners by enhancing their knowledge in connection with real-life situations while they train in mandatory safe conditions.*

**Introduction**

Haptic devices generate small forces through a mechanical linkage (e.g., a stylus in the user's hand), allowing the user to sense the shape and some material properties of virtual objects. Haptic hardware and associated technology have become increasingly more available, especially in entertainment (e.g. electronic games) and the medical field (e.g. simulation and training of surgical procedures) (Basdogan et al., 2004). In the area of medical diagnosis and minimally invasive surgery (e.g. laparoscopy) there is a strong need to determine mechanical properties of biological tissue for both histological and pathological considerations. One of the established diagnosis procedures is the palpation of body organs and tissue.

In this paper we present a visuo-haptic simulator designed to improve practice-based education in laparoscopy. We focus on liver palpation simulation and laparoscopic tools manipulation. The simulator can be used as a preliminary step for minimally invasive surgical training in liver related surgical procedures.

The paper is structured as follows. Section 2 presents a few facts about liver pathology as well as related work in laparoscopy simulation for liver based procedures. Section 3 presents the graphical and haptic user interface for the system. Section 4 presents the simulation cases of different liver pathologies followed by the assessment of the simulator in Section 5.

**Simulation and Training for Liver-based Laparoscopy Procedures**

The largest organ in the human body, the liver is also one of the most affected by disease. For example hepatitis C virus infection is a growing public health concern. Globally an estimated 180 million people, or roughly 3% of the world's population, are currently infected (Ford et al., 2012). The normal liver is smooth, with no irregularities. The smoothness is due the fact that the liver is covered in the most part by visceral peritoneum that forms its serous membrane. The liver has greater consistency than other glandular organs. It is tough and its percussion gives dullness. It is brittle and less elastic, so that it breaks and crushes easily. The liver has a high plasticity, which allows it to mould after neighbouring organs (Târcoveanu et al., 2005).

In minimally invasive surgery internal tissue palpation is an important pre-operatory activity (Khaled et al., 2004). Liver palpation can reveal multiple issues: presence of emphysema with an associated depressed diaphragm, fatty infiltration (enlarged with rounded edge), active hepatitis (enlarged and tender), cirrhosis (enlarged with nodular irregularity), hepatic neoplasm (nodular consistency).

State-of-art hepatic laparoscopic simulations, like other cutting-edge surgical simulations, take advantage of increased computational power and haptic device accuracy to supplement the pre-operative planning process and surgeons training.

The EU PASSPORT project is an example of a current laparoscopic liver resection simulation. The project utilizes "advanced methods and the computational power of today GPUs to simulate multiple organs with high-resolution deformations and collisions in real-time" (Passport, 2012). A similar research effort (Acharya et al., 2008) studied the effects of surrounding organ kinematics and geometry on liver access. The group modelled respiratory diaphragm motion for integration into surgical training and planning simulators. Villard (Villard et al., 2009) went a step further, including rib cage respiratory movement, soft tissue behaviour, and a collection of virtual patients and their organs, segmented from CT scans of actual patients in their liver biopsy simulator. These forward strides have necessitated parallel advances in the area of organ modelling.

Lister (Lister et al., 2011) developed a nonlinear liver model through experimental setups designed to collect precise measurements in force-displacement, surface deformation, and organ boundary conditions. The model was augmented with an outer capsule that constrained surface tissue movement for added realism. Model accuracy was assessed through a probing simulation. Beyond organ modelling, surgical procedure modelling has also improved. Marciel (Maciel at al., 2008) developed a real time physics-based virtual electrosurgical simulation tool in which heat generation in the tissue is linked to the applied electric potential. Such electro-surgery tasks are indispensable in laparoscopic surgery simulation specifically for a virtual liver ablation.

While 3D organ models have progressed in the last decade from linear (Delingette, 2000) to nonlinear (Ayache et al., 2003), simulations have grown increasingly complex and layered—imparting invaluable physiological knowledge and experience that may be otherwise impossible to attain.

**HapticMed Simulator**

During our business analysis phase, through discussions with surgeons from Constanta Regional Hospital we identified four scenarios for training in the HapticMed simulator. The first case presents a 3D haptic model of a healthy liver tissue; the second case focuses on the pathologic case of cirrhosis; the third case, on a liver with tumours and case number four simulates a hepatic liver.

**Hardware Components**

The main hardware components of our simulation system are: a set of two Phantom Omni (Sensable, 2012) devices and a 3D visualization system based on shutter glasses. A Maryland pense (see Figure 1) is attached to the Omni device and is restricted through a metal ring that simulates the trocar entry point.

**The User Interface**

The simulator allows users to interact with the virtual environment through a standard keyboard and one or two haptic devices simultaneously. The graphical interface consists of a set of 3D elements such as buttons, as well as 2D labels and text.

It is essential that the user familiarizes with the haptic device manipulation in a 3D virtual space before using the simulator. Therefore, the user must touch with the haptic device a sphere randomly positioned on the screen several times in a fixed time interval (see Figure 2). If the user does not succeed, s/he can retry the task several times until s/he becomes accustomed with the visuo-haptic interface.

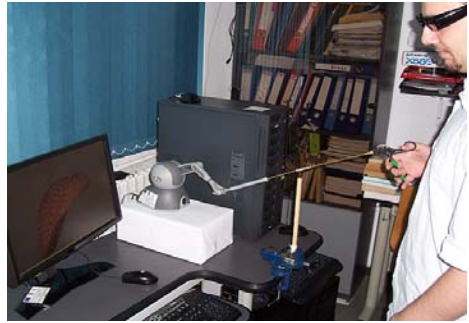

Figure 1. **Hardware components**

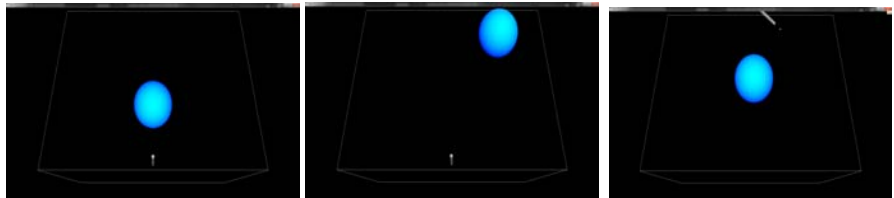

Figure 2. **Basic user-interface interaction for familiarization with the haptic interface**

**Simulated Scenarios**

**Healthy Liver and Cirrhotic Liver**

The first interaction between the user and the interface is on a healthy liver model. A 3D deformable model of the liver is presented to the user and the interaction is possible though a Maryland pense as well as a Babcock pense which is broad, has flared ends with smooth tips allowing tissue palpation. All current scenarios assumes that the laparoscopic camera and the corresponding light source are fixed and do not require user attention. The visual-haptic interface is presented in Figure 3.

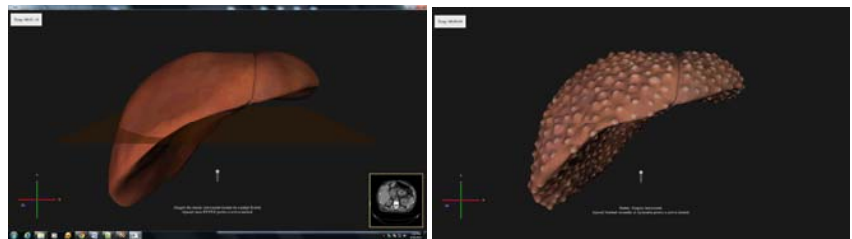

a) Generic Interface

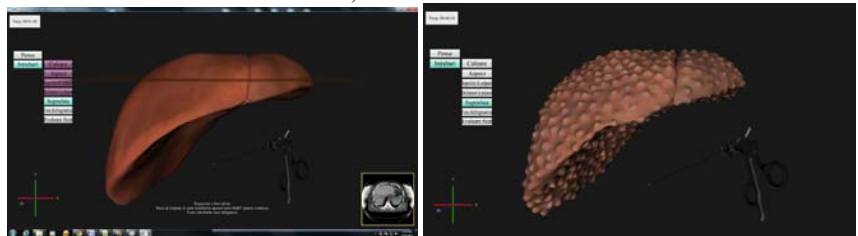

b) Menu on the left for choosing questions

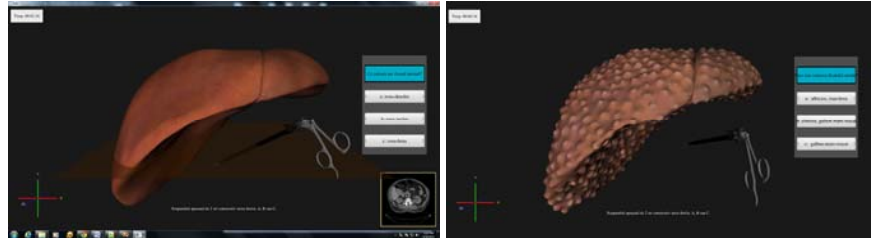

c) Menu on the right for selecting answers

Figure 3. **HapticMed simulator work session.**
Healty liver evaluation (left), cirrhotic liver evaluation (right).

The goal for the Healthy Liver scenario is to complement the theoretical knowledge of the student by allowing him to palpate and obtain realistic force feedback from a healthy liver tissue. The improvement and evaluation processes for liver palpation focuses on the force range (min-max) applied during palpation, the direction of force application (based on the instrument angle to the surface) as well as the palpation methodology and palpation zones/areas.

In the Cirrhotic Liver scenario the user uses the pense to explore through touch the liver surface properties. After palpating the surface bumps, observing their consistency and frequency (Figure 3 – right side images), the user employs a menu system to present the disease condition based on attributes like tissue color and consistency.

**Liver Tumours and Hepatic Liver Scenarios**

These two scenarios follow the same evaluation structure like the ones for the healthy and cirrhotic liver: choosing questions, palpation execution and question answering. The tumor model presents two types of cysts: one type is visible at the liver surface and presents stiffness properties different for the rest of the liver surface, the other one is internal cysts (deep cysts) that are not visible at the surface however can be detected haptically through surface palpation. A successful liver evaluation in this case requires a full surface palpation to identify surface as well as potential deep cysts.

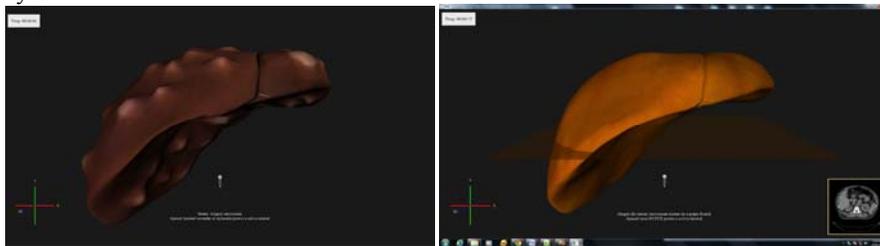

a)　　　Generic Interface

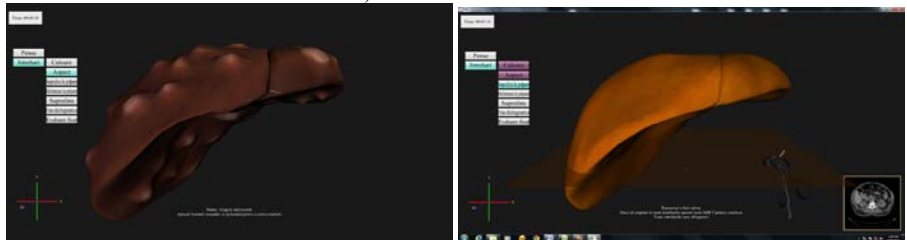

b) Menu on the left for choosing questions

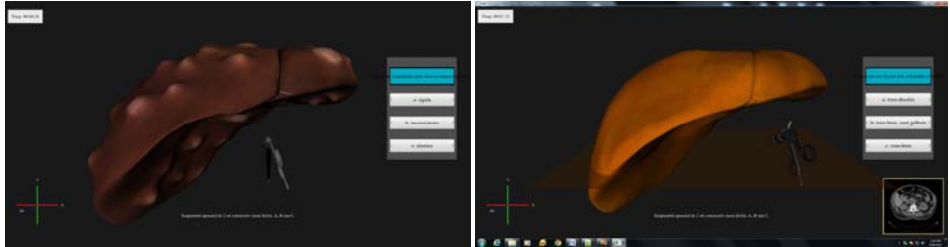

c) Menu on the right for selecting answers
Figure 4. **HapticMed simulator work session**
Liver with tumors/cysts evaluation (left), Hepatic liver evaluation (right).

The hepatic liver simulation presents a visually as well as haptically modified liver model. In comparison with the healthy liver, the hepatic liver surface color is more pale and the tissue consistency is significantly increased.

**Simulator Assessment**

The force applied during palpation must be maintained in a certain range. Palpation with small forces may not reveal correctly mechanical properties of the biological tissue, while forces exceeding a certain threshold can irreversibly damage healthy liver tissue.

**Interactive Palpation Force Measurement**

We proposed and implemented a dynamic force measurement approach and visualization module to find the appropriate range of forces during the liver palpation procedure, collecting force data directly from the experienced surgeons we cooperate with. The module draws a force measurement indicator range on the left side of the screen as illustrated in Figure 5.

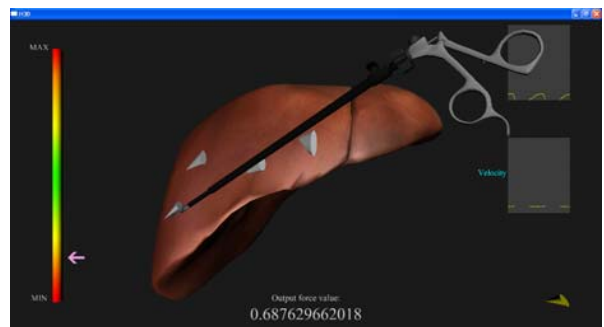

Figure 5**. Dynamic force measurement and display**

The range empirically agreed upon is in the interval 2.1 to 2.5 Newtons. A standard Babcock pense was connected to the haptic device and used to practice palpation.

**Force Map Visualization**

The prototype we developed represents the palpation force, position and orientation thought cones directly on the liver's surface. The cone's height and bottom radius are proportional with the magnitude of the force applied on the tissue's surface. Moreover the position and orientation of the pense is represented by the cone's height direction. So the evaluator can see not only the force applied but also the location and the direction of the pense relative to the liver surface. The assessment method takes into consideration the palpation gesture according to the type of liver the user evaluates: the recommended palpation force used to a normal liver differs from the one used on a hepatic liver.

In Figure 6 (left) the user is an experienced surgeon: the palpation force used on each "tap" on the liver's surface is constant. We observe that the velocity of the Babcock pense on the liver surface is constant too. In Figure 6 (right) the user is a novice: the palpation force and the haptic device's velocity vary abruptly when it should remain at a relative constant value to avoid tissue damage.

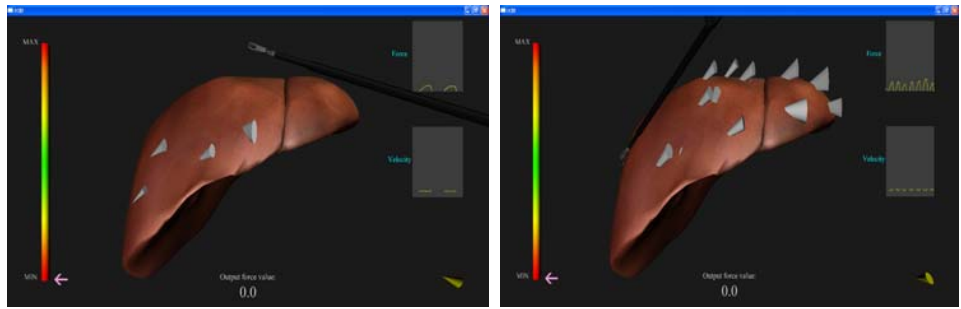

Figure 6. **Force map visualization (experienced-left, novice-right)**

**Conclusions**

Haptic devices, allowing the simulation of touch are becoming increasingly available and affordable. Their use in medical simulation and training has been recognized worldwide for more than a decade.

We have developed the first 3D visual and haptic simulator for liver diagnostic through palpation in Romania. This custom built simulator has enabled development of new expertise in haptic system development and integration for Romanian computer science and engineering students. As opposed to commercial simulators for laparoscopic procedures, our simulator is a fraction of the cost and has been developed mainly with open source software. The results obtained so far point to direct applications in the medical industry and practice. The simulator can improve medical training thus helping save human lives.

We are in the process of assessing the simulator by the surgeon residents from the Regional Hospital of Constanta, Romania as well as developing new research collaborations with universities and research groups from Europe and US.

**Acknowledgments**
This study was supported under the ANCS Grant "HapticMed – Using haptic interfaces in medical applications", no. 128/02.06.2010, ID/SMIS 567/12271, POSCCE O.2.1.2 / 2009.